\documentclass[aps,prl,superscriptaddress,showpacs,floatfix,nofootinbib,notitlepage,twocolumn]{revtex4-2}
\usepackage{amsmath,graphicx,float,hyperref,csquotes,mdframed,appendix}

\def\bra{\langle}
\def\ket{\rangle}

\newcommand{\trento}{T$\mathrel{\protect\raisebox{-2.1pt}{R}}$ENTo}

\newcommand{\pbpb}{$^{208}$Pb+$^{208}$Pb}

\newcommand{\uuuu}{$^{238}$U+$^{238}$U}
\newcommand{\xexe}{$^{129}$Xe+$^{129}$Xe}

\begin{document}

\title{Evidence of the triaxial structure of $\boldsymbol{^{129}}$Xe at the Large Hadron Collider}

\author{Benjamin Bally}
\affiliation{Departamento de Física Teórica, Universidad Autónoma de Madrid, 28049 Madrid, Spain}

\author{Michael Bender}
\affiliation{Université de Lyon, Institut de Physique des 2 Infinis de Lyon,
IN2P3-CNRS-UCBL, 4 rue Enrico Fermi, 69622 Villeurbanne, France}

\author{Giuliano Giacalone}
\affiliation{Institut f\"ur Theoretische Physik, Universit\"at Heidelberg,
Philosophenweg 16, 69120 Heidelberg, Germany}

\author{Vittorio Som\`a}
\affiliation{IRFU, CEA, Universit\'e Paris-Saclay, 91191 Gif-sur-Yvette, France}

\begin{abstract}
    The interpretation of the emergent collective behaviour of atomic nuclei in terms of deformed intrinsic shapes  \cite{BohrMottelson} is at the heart of our understanding of the rich phenomenology of their structure, ranging from nuclear energy to astrophysical applications across a vast spectrum of energy scales. A new window onto the deformation of nuclei has been recently opened with the realization that nuclear collision experiments performed at high-energy colliders, such as the CERN Large Hadron Collider (LHC), enable experimenters to identify the relative orientation of the colliding ions in a way that magnifies the manifestations of their intrinsic deformation \cite{Giacalone:2019pca}. Here we apply this technique to LHC data on collisions of $^{129}$Xe nuclei \cite{ALICE:2018lao,CMS:2019cyz,ATLAS:2019dct} to exhibit the first evidence of non-axiality in the ground state of ions collided at high energy. We predict that the low-energy structure of $^{129}$Xe is triaxial (a spheroid with three unequal axes), and show that such deformation can be determined from high-energy data. This result demonstrates the unique capabilities of precision collider machines such as the LHC as new means to perform imaging of the collective structure of atomic nuclei.
\end{abstract}

\maketitle

\begin{figure}[t]
    \centering
    \includegraphics[width=.75\linewidth]{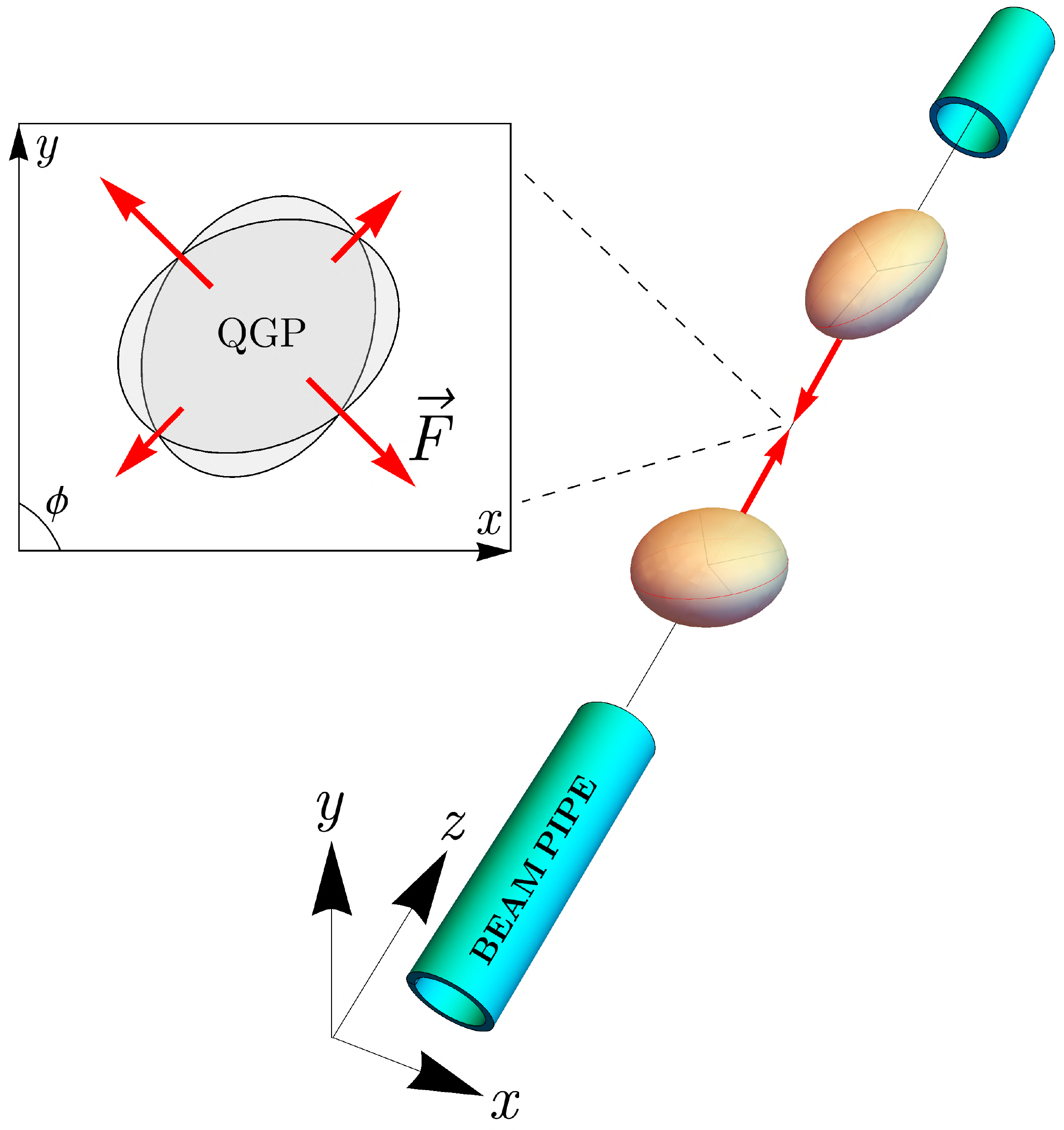}
    \caption{Illustration of a head-on collision between two atomic nuclei performed in a collider experiment. The nuclei, deformed in their ground state, are randomly oriented as they run in the beam pipe, and the shape of their area of overlap can range from circular to elliptical. A quark-gluon plasma (QGP) is formed in the this area. The hydrodynamic expansion of this medium in the plane transverse to the beam, $(x,y)$, is driven by a force field, $\vec F$, which carries the same quadrupole anisotropy as the QGP geometry, i.e., as the overlap area. Note that in the frame of the laboratory both nuclei would look like thin pancakes, squeezed in beam direction, $z$, due to a strong effect of Lorentz contraction. }
    \label{fig:1}
\end{figure}

 A key signature of the formation of quark-gluon plasma (QGP \cite{Bernhard:2019bmu,Gardim:2019xjs}) in nuclear collision experiments performed at high-energy colliders is the observation of sizable angular anisotropy in the emission of hadrons in the plane orthogonal to the collision axis (dubbed \textit{transverse plane}, or $(x,y)$ in Fig.~\ref{fig:1}). If $N$ hadrons are detected in a given collision event, their transverse angular distribution, $dN/d\phi$, where $\phi$ is the azimuthal angle, presents a quadrupole (elliptical) component \cite{Heinz:2013th}: 
\begin{equation}
    dN/d\phi \propto 1 + 2 v_2 \cos \bigl (2(\phi-\phi_2) \bigr )
\end{equation}
where $v_2$, dubbed elliptic flow, is the magnitude of the quadrupole asymmetry. $v_2$ is engendered in nuclear collisions by the pressure gradient force, $\vec F=-\vec \nabla P$, driving the QGP expansion that converts the spatial anisotropy of the system geometry, which has in general some ellipticity \cite{Ollitrault:1992bk,PHOBOS:2006dbo}, into an anisotropic flow of matter, carried over to the detected hadrons.
\begin{figure*}[t]
    \centering
    \includegraphics[width=.75\linewidth]{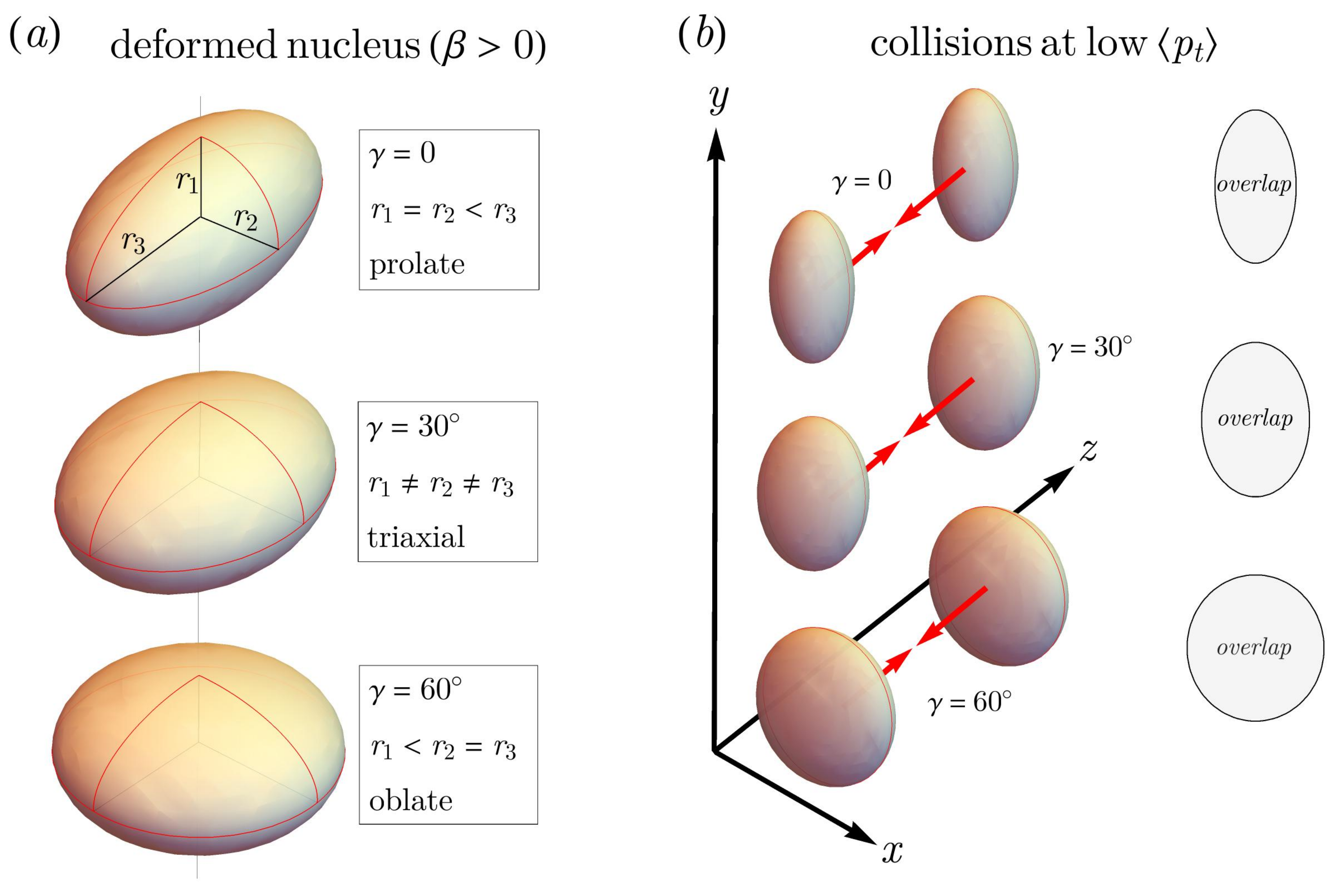}
    \caption{Meaning of the triaxial parameter, $\gamma$, and its influence on the geometry of head-on collisions of nuclei at low mean transverse momentum, $\bra p_t \ket$. (a) Spheroids with a quadrupole deformation, $\beta>0$. Depending on the value of $\gamma$, they can either present two axes of the same length, and be either prolate ($\gamma=0$) or oblate ($\gamma=60^{\circ}$), or present three axes of different lengths and be triaxial ($0<\gamma<60^\circ$), with the maximum triaxiality being reached for $\gamma=30^\circ$. Values of $\gamma$ between $60^\circ$ and $360^\circ$ correspond simply to rotations of the same types of shapes. (b) Head-on collisions at small values of the mean transverse momentum of detected hadrons, $\bra p_t \ket$, permit to isolate configurations that maximize the overlap area. Depending on the value of $\gamma$, collision configurations at low $\bra p_t \ket$ can thus vary between geometries that maximally break azimuthal symmetry ($\gamma=0$) and geometries that are azimuthally symmetric ($\gamma=60^\circ$).}
    \label{fig:2}
\end{figure*}
For nearly head-on (\textit{central}) collisions (Fig.~\ref{fig:1}), elliptic flow is naturally sensitive to the intrinsic quadrupole deformation that characterizes the ground state of colliding ions, i.e., the ellipsoidal deformation of their surface,
\begin{equation}
\label{eq:1}
    R(\theta,\varphi) = R_0 \left\{ 1 + \beta \left[  \cos \gamma Y_{20}(\theta,\varphi) + \sin \gamma Y_{22}(\theta,\varphi) \right] \right\},
\end{equation}
where $R_0$ parametrizes the nuclear radius, the $Y_{lm}$ are spherical harmonics, and the positive coefficients $\beta$ and $\gamma$ encode the ellipsoidal shape. The former gives the magnitude of deformation, with well-deformed nuclei having $\beta\approx0.3$, while the latter indicates the length imbalance of the axes of the spheroid (e.g., if the nucleus is prolate, like a rugby ball, or oblate, flattened at the poles), and varies between $\gamma=0$ and $\gamma=60^\circ$, as shown by Fig.~\ref{fig:2}(a).

Spectroscopy of nuclei at low energy provides access to the nuclear charge quadrupole moment that can also be parameterized with coefficients $\beta_v$ and $\gamma_v$ of similar size as (albeit not equivalent to \cite{Hasse:1988}) $\beta$ and $\gamma$ in Eq.~(\ref{eq:1}). For well-deformed nuclei, $\beta_v$ can be determined through the measurement of a single transition probability \cite{Pritychenko:2013gwa}, whereas access to several transitions is required for $\gamma_v$ \cite{Morrison:2020}. Identifying these parameters with a nuclear shape requires assuming that the nuclear wave function can be factorized into an intrinsic state describing the geometrical arrangement of the nucleons, and a state giving the orientation of this structure in the laboratory frame \cite{BohrMottelson,Bally:2021}. Such factorization is not always meaningful, for example because of large shape fluctuations. Additionally, for odd-mass and odd-odd nuclei, quantum mechanics entangles the contributions from collective intrinsic deformation and individual single-particle states to the quadrupole moment, such that the former cannot be uniquely determined from a spectroscopic experiment. 

Measurements of $v_2$ at colliders provide an alternative access route to the intrinsic deformation of all nuclei, even and odd. The impact of $\beta_v$ at high energy has been established in \uuuu{} collisions at the Relativistic Heavy Ion Collider (RHIC) \cite{STAR:2015mki}, and in \xexe{} collisions at the LHC \cite{ALICE:2018lao,CMS:2019cyz,ATLAS:2019dct}. Our goal is to show that such experiments open as well a new window onto $\gamma_v$. The key feature is the possibility of selecting events for which the relative orientation of the colliding ions maximizes the breaking of azimuthal symmetry induced by their shapes \cite{Giacalone:2019pca}. One needs the mean hadron momentum,
\begin{equation}
\label{eq:mpt}
    \bra p_t \ket = \frac{1}{N} \sum_{i=1}^N p_{t,i},
\end{equation}
where $p_{t,i}\equiv|{\bf p}_{t,i}|$ for particle $i$ with transverse momentum ${\bf p}_{t,i}=(p_{x,i},p_{y,i})$. For collisions at fixed $N$, $\bra p_t \ket$ provides a measure of the (inverse) size of the transverse area where the QGP is formed \cite{Broniowski:2009fm}, such that events carrying abnormally small values of $\bra p_t \ket$ correspond to large overlap areas. Following Fig.~\ref{fig:2}(b), when the ions have $\beta>0$, low-$\bra p_t \ket$ configurations correspond to overlap geometries ranging from maximally elliptic, for $\gamma=0$, to azimuthally isotropic, for $\gamma=60^\circ$. At low $\bra p_t \ket$, then, the magnitude of $v_2$ depends on $\gamma$, so that the dependence of $v_2$ on $\bra p_t \ket$ probes $\gamma$. For strongly prolate $^{238}$U nuclei with $\beta\approx0.3$ and $\gamma\approx0$, the effectiveness of this method in probing $\beta$ has been recently demonstrated in \uuuu{} collisions \cite{jia}. Here we employ this technique to reveal for the first time signatures of $\gamma$, the triaxiality of nuclei. We use LHC measurements in \xexe{} collisions, which are ideal candidates for such a study, as the ground state of all even-mass xenon isotopes around $^{129}$Xe are triaxial in low-energy nuclear models \cite{Delaroche:2009fa,Scamps:2021}.

\begin{figure}[t]
    \centering
    \includegraphics[width=.95\linewidth]{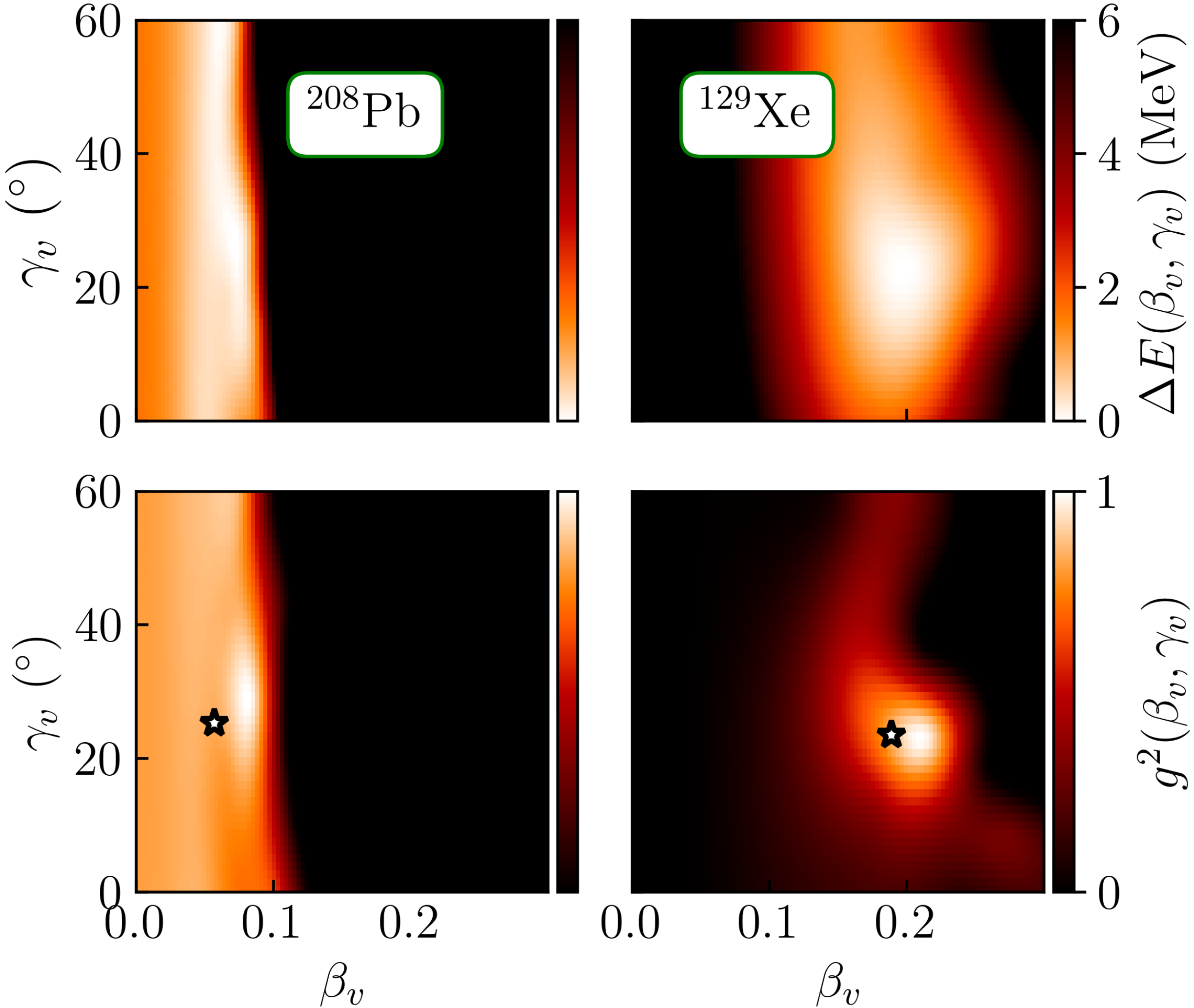}
    \caption{Structure of the ground states of $^{129}$Xe and $^{208}$Pb. Left: $^{208}$Pb. Right: $^{129}$Xe. The upper panels represent beyond-mean-field potential energy surfaces in the $(\beta_v,\gamma_v)$ plane, where we plot the energy shift, $\Delta E$, with respect to the energy minimum. The lower panels show the functions $g^2(\beta_v,\gamma_v)$ normalized to unity at their maximum. Star markers label the coordinates of the average intrinsic quadrupole moments for both nuclei, namely,  $\bar \beta_v=0.06$, $\bar \gamma_v=25.3^\circ$ for $^{208}$Pb, and $\bar \beta_v=0.19$, $\bar \gamma_v=23.6^\circ$ for $^{129}$Xe. We note that the white regions in the displayed color maps correspond to the minimum of $\Delta E(\beta_v,\gamma_v)$ in the upper panels, and to the maximum of $g^2(\beta_v,\gamma_v)$ in the lower panels.}
    \label{fig:3}
\end{figure}

We determine now in the framework of energy-density functional methods \cite{Bender03a} applied to the nuclear many-body problem the triaxiality of the lowest $1/2^+$ state of $^{129}$Xe, corresponding to the experimental ground state. We first perform a set of symmetry-breaking constrained Hartree-Fock-Bogoliubov (HFB) calculations that provide intrinsic states covering the $(\beta_v, \gamma_v)$ plane. The lowest state with good quantum numbers in the laboratory frame is then constructed using the projected generator coordinate method (PGCM), i.e., we consider a many-body wave function that is a linear superposition of the intrinsic states across the $(\beta_v, \gamma_v)$ plane, projected onto quantum numbers reflecting the symmetries of the nuclear Hamiltonian. From there, we compute the so-called collective wave function $g(\beta_v,\gamma_v)$ \cite{Bender03a}, which squared gives roughly the contribution of each $(\beta_v,\gamma_v)$ point to the final PGCM state. The same effective Skyrme-type nucleon-nucleon interaction, SLyMR1 \cite{Sadoudi:2013,JodonPHD}, is used at all stages of the calculations, done for $^{129}$Xe and $^{208}$Pb, as we shall look at high-energy results for both these species. The structural properties of these nuclei are shown in Fig.~\ref{fig:3}. The results for $^{208}$Pb agree with existing literature \cite{Delaroche:2009fa}, indicating a soft energy surface, with all states being nearly degenerate irrespective of their $\gamma_v$, up to $\beta_v\approx0.1$, beyond which the energy rises quickly. Our new result concerning $^{129}$Xe shows instead a minimum around $\beta_v=0.2$, corresponding to a $g^2(\beta_v,\gamma_v)$ peaked around the average intrinsic moments $(\bar \beta_v, \bar \gamma_v) = (0.19,23.6^\circ)$. $^{129}$Xe appears to be, hence, a rigid triaxial spheroid.

\begin{figure*}[t]
    \centering
    \includegraphics[width=.85\linewidth]{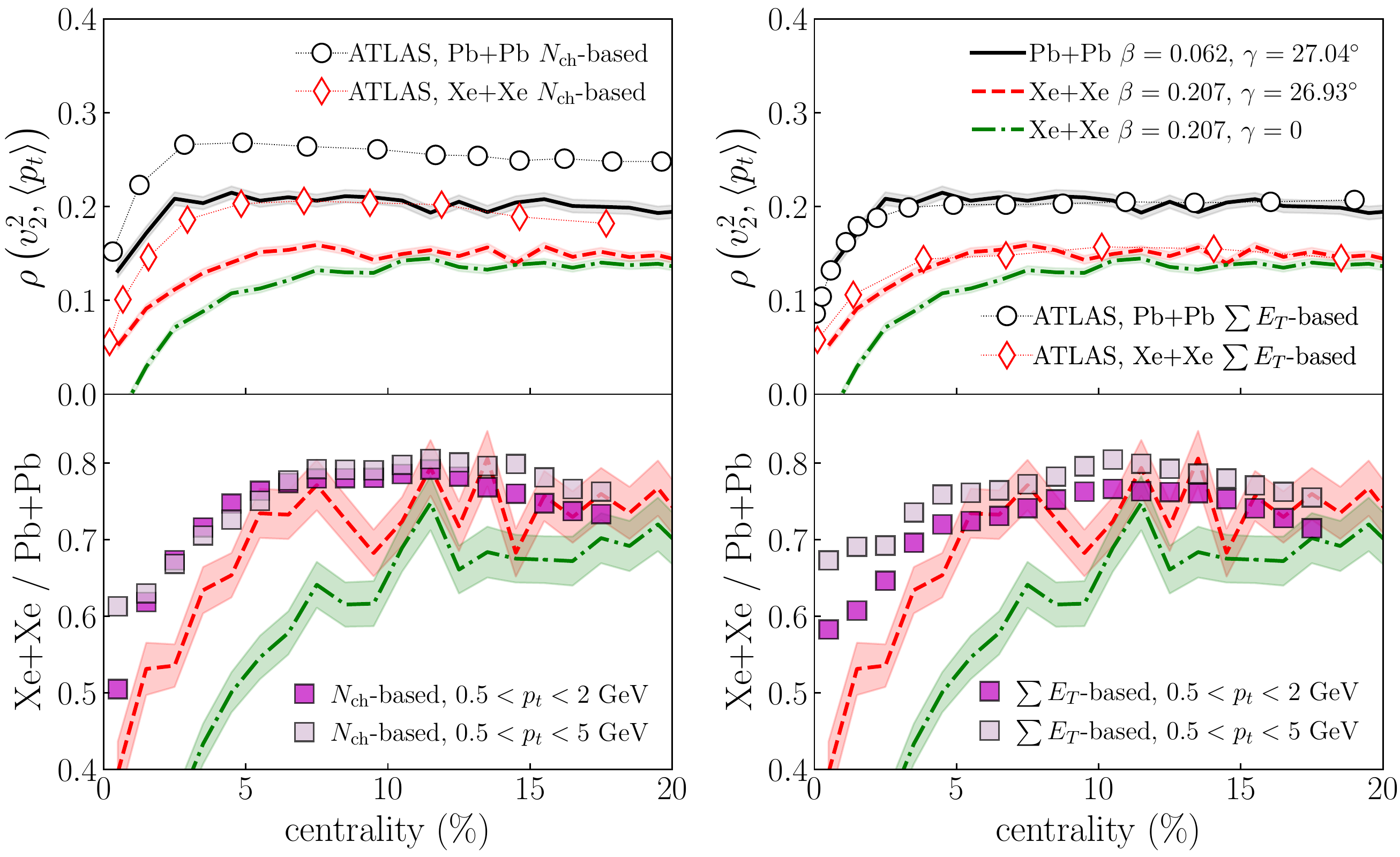}
    \caption{Theoretical and experimental results on the correlation between elliptic flow and the mean transverse momentum.  Symbols are preliminary measurements from the ATLAS collaboration (circles, diamonds, and squares). Error bars on ATLAS data are of the same size as the displayed symbols. Theory results include collisions of $^{208}$Pb (black solid line), collisions of triaxial $^{129}$Xe nuclei (red dashed lines), as well as of $^{129}$Xe nuclei with $\gamma=0$ (green dot-dashed lines). The upper panels show the centrality dependence of the correlator $\rho(v_2^2,\bra p_t \ket)$ in both \pbpb{} and \xexe{} collisions. The lower panels show their ratio. The reconstruction of the impact parameter in ATLAS data is based on either a raw number of charged hadrons, $N_{\rm ch}$ (left panels), or the energy collected by dedicated calorimeters, $\sum E_T$ (right panels). The data points in the upper panels are obtained from hadrons having $0.5<p_t<2$ GeV, while the ratios in the lower panels are calculated as well for $0.5<p_t<5$ GeV \cite{ATLAS:2021kty}. The shaded bands represent the statistical uncertainties on the theoretical results.}
    \label{fig:4}
\end{figure*}

With this knowledge, we perform simulations of \pbpb{} and \xexe{} collisions to assess the role of $\gamma$ in high-energy experiments. Following the Glauber Monte Carlo model \cite{Miller:2007ri}, the colliding nuclei are treated as batches of independent nucleons sampled from a density, $n$, usually taken as a Woods-Saxon profile:
\begin{equation}
\label{eq:ws}
n(r,\theta,\varphi) \propto \biggl ( 1 + \exp\biggl [ \frac{1}{a} \biggl (r-R(\theta,\varphi) \biggr ) \biggr ] \biggr  )^{-1},   
\end{equation}
where $a$ is the skin thickness, and $R(\theta,\varphi)$ has the same form as in Eq.~(\ref{eq:1}). The parameters entering Eq.~(\ref{eq:ws}) are obtained by fitting the Woods-Saxon profile to the one-body nucleon density returned by a HFB calculation with SLyMR1, constrained to present quadrupole moments $\bar \beta_v$ and $\bar \gamma_v$. The fit yields: $a=0.537$~fm, $R_0=6.647$~fm, $\beta=0.062$, $\gamma=27.04^\circ$ for $^{208}$Pb, and $a=0.492$~fm, $R_0=5.601$~fm, $\beta=0.207$, $\gamma=26.93^\circ$ for $^{129}$Xe. Supplementing the Glauber model with an ansatz \cite{Moreland:2014oya} for the energy density of the QGP created in each collision, we study the impact of $\gamma$ on the dependence of $v_2$ on $\bra p_t \ket$ by evaluating the Pearson correlation coefficient \cite{Bozek:2016yoj}:
\begin{equation}
\label{eq:5}
    \rho(v_2^2,\bra p_t \ket) = \frac{\bigl  \bra \delta v_2^2 \delta \bra p_t \ket \bigr \ket}{\sqrt{\bigl \bra \bigl (\delta v_2^2 \bigr)^2 \bigr \ket  \bigl \bra \bigl (\delta \bra p_t \ket \bigr)^2  \bigr \ket}},
\end{equation}
where $\bra \ldots \ket$ denotes an average over events at fixed centrality, and $\delta o = o - \bra o \ket$ for any observable $o$. This quantity is a number between -1 (perfect anti-correlation) and +1 (perfect correlation). In central collisions, one expects $\rho(v_2^2,\bra p_t \ket)>0$ \cite{ATLAS:2019pvn}. Figure~\ref{fig:2} shows that, for prolate nuclei, decreasing the value of $\bra p_t \ket$ yields maximally elliptical overlap geometries, leading to enhanced values of $v_2$. Therefore, for $\beta>0$ and $\gamma=0$, nuclear deformation yields a negative contribution to $\rho(v_2^2,\bra p_t \ket)$, which gradually turns into a positive one towards $\gamma=60^\circ$ \cite{Jia:2021wbq}.

We evaluate Eq.~(\ref{eq:5}) as a function of the percentage of overlap (or \textit{impact parameter}) of the colliding ions, represented as a percentile, where 20\% corresponds roughly to a distance between the colliding ions of 7 (6) femtometers for \pbpb{} (\xexe{}) events. Our results are in Fig.~\ref{fig:4}. In central collisions, the role of $\gamma$ is manifest. Colliding triaxial $^{129}$Xe nuclei (red dashed line) enhances $\rho(v_2^2, \bra p_t \ket)$ compared to the case where the nuclei have $\gamma=0$ (green dot-dashed line), flipping its sign for the lowest percentiles. This sensitivity of $\rho(v_2^2, \bra p_t \ket)$ to the triaxiality implies that dedicated simulation frameworks \cite{Nijs:2020ors,JETSCAPE:2020shq} will be able to perform independent extractions of $\gamma$ from high-energy data. Comparing our results with preliminary LHC measurements by the ATLAS collaboration \cite{ATLAS:2021kty}, shown in the upper panels of Fig.~\ref{fig:4}, we see that they capture the trend of the data, although the effect of the experimental uncertainty on the centrality definition, causing the differences between the data points displayed in the left and in the right panel of the figure, is not included in our calculation. However, such effects become less relevant in the ratio of the two collision systems, shown in the lower panels (purple squares). The $p_t$ range of the hadrons used in the analysis, while influencing $\rho(v_2^2, \bra p_t \ket)$ \cite{ATLAS:2021kty}, plays also a minor role in Fig.~\ref{fig:4}. The measured ratios are, hence, very robust. They grant access to the triaxiality from data, and confirm our low-energy prediction that $^{129}$Xe has $\beta\approx0.2$ and $\gamma\approx27^\circ$. A similar analysis can be repeated for any species used in collider experiments, for instance, $^{197}$Au nuclei collided at RHIC \cite{jia}, which may also present a triaxial ground state \cite{Scamps:2021}. The possibility of knowing $\gamma$ from high-energy data represents a stepping stone to a fruitful collaboration between the low- and the high-energy nuclear communities, to better exploit collider experiments involving atomic nuclei.

\bigskip
  This project has received funding from the European Union’s Horizon 2020 research and innovation programme under the Marie Skłodowska-Curie grant agreement No. 839847. 
  M.B. acknowledges support by the French Agence Nationale de la Recherche under grant No. 19-CE31-0015-01 (NEWFUN).
  G.G. is supported by the Deutsche Forschungsgemeinschaft (DFG, German Research Foundation) under Germany’s Excellence Strategy EXC 2181/1 - 390900948 (the Heidelberg STRUCTURES Excellence Cluster), SFB 1225 (ISOQUANT) and FL 736/3-1. We acknowledge the computer resources and assistance provided by the Centro de Computación Científica-Universidad Autónoma de Madrid (CCC-UAM).

\appendix

\section{Supplemental material}

The calculations we have performed to produce the results presented in this manuscript are of two kinds. 
\begin{enumerate}
    \item We first evaluate the structure properties of the first 1/2$^+$ state of $^{129}$Xe and the first $0^+$ state of $^{208}$Pb. The goal is to extract the average deformation parameters $\beta$ and $\gamma$ of these species, as well as the parameterization of their radial profiles.
    \item Secondly, we use the knowledge of the structural properties of these nuclei to perform Monte Carlo simulations of millions of high-energy \xexe{} and \pbpb{} collisions to produce the results shown in Fig.~\ref{fig:4}.
\end{enumerate}

\bigskip
\noindent 1.~Within our nuclear structure model, we build approximations to the eigenstates of the nuclear Hamiltonian $H$ by solving the variational equation
\begin{equation}
  \delta \frac{\langle \Psi \vert H \vert \Psi \rangle}{\langle \Psi \vert \Psi \rangle} = 0 
\end{equation}
in a restricted many-body Hilbert space. Here, the nuclear Hamiltonian is taken to be the SLyMR1 parametrization of a phenomenological Skyrme-type pseudopotential \cite{Sadoudi:2013,JodonPHD}.

In the first step of the calculation, the variational equation is solved considering Bogoliubov quasiparticle states as trial wave functions \cite{Bender03a}. A series of constrained minimizations of the energy is performed to explore the $(\beta_v,\gamma_v)$ surface and we end up with a set of Bogoliubov-type wave functions $\left\{ \vert \Phi(\beta_v, \gamma_v) \rangle \right\}$ that are characterized by their average dimensionless quadrupole moments
\begin{align}
\label{eq:beta:V}
\beta_v &= \frac{4\pi}{(3 R_0^{2} A)} \sqrt{q_{20}^2 + 2 q_{22}^2} , \\
\label{eq:gamma:V}
\gamma_v &= \arctan\left(\frac{\sqrt{2} q_{22}}{q_{20}} \right) , 
\end{align}
where $A$ is the mass number, $R_0=1.2A^{1/3}$, and
\begin{align}
Q_{lm} &\equiv r^{l} Y_{lm}(\theta,\varphi) , \\
q_{lm} & \equiv \frac12 \langle \Phi(\beta_v,\gamma_v) \vert Q_{lm} + (-1)^m   Q_{l -m}  \vert \Phi(\beta_v,\gamma_v) \rangle .
\end{align}
Then, in the second step of the calculation, we consider an enriched variational ansatz that is built as a linear superposition of the states in the set $\left\{ \vert \Phi(\beta_v,\gamma_v) \rangle \right\}$ projected onto good quantum numbers associated with the symmetries of the nuclear Hamiltonian \cite{Bally:2021}. This is the PGCM that was used for example in \cite{Bally14a}. The variational ansatz now reads as 
\begin{equation}
\label{eq:collective:state}
 \vert \Psi \rangle = \sum_{(\beta_v,\gamma_v) K} f_{(\beta_v,\gamma_v) K} P^{J}_{MK} P^{N} P^{Z} |\Phi(\beta_v,\gamma_v)\rangle ,
\end{equation}
where $P^{J}_{MK}$, $P^{N}$, $P^{Z}$ are projection operators onto good angular momentum $J$ and its projections along the $z$ axis $M$, $K$, neutron number $N$ and proton number $Z$, respectively \cite{Bally:2021}. The weights $f_{(\beta_v,\gamma_v) K}$ as well as the energy of the state $\vert \Psi \rangle$ are solutions of the variational equation, which is in this case is equivalent to a generalized eigenvalue problem. In practice, the sets used to build the PGCM anzatz contained 15 states for $^{208}$Pb and 23 states for $^{129}$Xe. 
In particular, for $^{129}$Xe, the calculations were tailored to obtain the best possible description of the lowest $1/2^+$, which is the experimental ground state but is probably an excited state when performing a large-scale PGCM calculation of $^{129}$Xe with the SLyMR1 parametrization. 

After obtaining the correlated PGCM wave functions, we extract average quadrupole moments $(\bar{\beta}_v,\bar{\gamma}_v)$ for the state $\vert \Psi \rangle$ by computing
\begin{align}
 \bar{\beta}_v &= \sum_{(\beta_v,\gamma_v)} \beta_v \, g^2(\beta_v,\gamma_v) , \\
 \bar{\gamma}_v &= \sum_{(\beta_v,\gamma_v)} \gamma_v \,  g^2(\beta_v,\gamma_v) ,
\end{align}
where $g(\beta_v,\gamma_v)$ is the so-called collective wave function, whose expression in terms of the $f_{(\beta_v,\gamma_v) K}$ of Eq.~(\ref{eq:collective:state}) can be found in \cite{Bender03a}.
We obtain $(\bar{\beta}_v=0.06, \bar{\gamma}_v=25.3^\circ)$ for $^{208}$Pb and $(\bar{\beta}_v=0.19, \bar{\gamma}_v=23.6^\circ)$ for $^{129}$Xe. To give confidence in the fact that our calculations correctly capture the structure of the experimental levels, we mention in particular that, when including suitable corrections for the finite
size of nucleons \cite{Cipollone15a}, our PGCM wave functions give charge radii $r_{\text{ch}} = 5.46$ fm for $^{208}$Pb, and $r_{\text{ch}} = 4.74$ fm for $^{129}$Xe, that reproduce experimental data with an accuracy better than one percent.

The average multipole moments of the point-nucleon density are used to inform our high-energy collision model in the following way. First, for a given nucleus, we compute a new Bogoliubov-type state $\vert \Phi(\bar{\beta}_v, \bar{\gamma}_v) \rangle$ that is constrained to the appropriate average intrinsic multipole moment. Then, we fit a Woods-Saxon density 
\begin{equation}
\label{eq:M1}
n(r,\theta,\varphi) \propto \biggl ( 1 + \exp\biggl [ \frac{1}{a} \biggl (r-R(\theta,\varphi) \biggr ) \biggr ] \biggr  )^{-1},   
\end{equation}
with
\begin{equation}
\label{eq:M2}
    R(\theta,\varphi) = R_0 \bigl [1 + \beta \bigl (  \cos \gamma Y_{20}(\theta,\varphi) + \sin \gamma Y_{22}(\theta,\varphi) \bigr) \bigr ],
\end{equation}
such that it reproduces at the same time the one-body density and average quadrupole moments of the state $\vert \Phi(\bar{\beta}_v, \bar{\gamma}_v) \rangle$. With this procedure, we obtain the parameters are $a=0.537$~fm, $R_0=6.647$~fm, $\beta=0.062$, $\gamma=27.04^\circ$ for $^{208}$Pb, and $a=0.492$~fm, $R_0=5.601$~fm, $\beta=0.207$, $\gamma=26.93^\circ$ for $^{129}$Xe. This Bogoliubov state is not only an approximation for the lowest state with good quantum numbers after symmetry restoration, but also incorporates effect of shape fluctuations, which for the nuclei discussed here is possible because their wave functions have only one peak. The Bogoliubov-type states $\vert \Phi(\beta_v, \gamma_v) \rangle$ have in general also higher-order multipole moments \cite{Scamps:2021}, whose presence influences the values for $\beta$ and $\gamma$. Including higher-order deformations in the shape of the Woods-Saxon density of Eq.~(\ref{eq:M2}), however, does not have a significant impact on the high-energy observables discussed here \cite{Jia:2021tzt}. But we note that the parameters $\beta$ and $\gamma$ of the fitted Woods-Saxon density are in general different from the average $(\bar{\beta}_v,\bar{\gamma}_v)$ of the microscopic wave function.

\bigskip
\noindent 2.~The simulations of \pbpb{} and \xexe{} collisions are subsequently performed within the framework of the Glauber Monte Carlo model \cite{Miller:2007ri}. 

The colliding nuclei are treated as batches of $A$ nucleons, which, in each realization of the nuclei, are sampled independently from the distribution of Eq.~(\ref{eq:M1}). Before sampling the nucleons, we consider that the spatial orientation of the colliding ions is random at the time of scattering. We randomly rotate the Woods-Saxon densities in space following the so-called Z-X-Z (or 3-1-3) prescription. Consider that the intrinsic nuclear frame and the lab frame, $(x,y,z)$, are initially aligned. We perform $i)$ a rotation about the $z$ axis by an angle $u$; $ii)$ a rotation about the $x$ axis by an angle $v$; $iii)$ an additional rotation about the $z$ axis by an angle $w$. To ensure that the rotations of the ellipsoids are sampled uniformly from SO(3), the angle $u$ an the angle $w$ are sampled uniformly between 0 and $2\pi$, whereas the angle $v$ is sampled such that the distribution of $\cos(v)$ is uniform between -1 and 1. Subsequently, we sample $A$ nucleons for each ion according to the rotated Woods-Saxon densities. We neglect any effect of short-range correlations in our nuclei, and do not impose, e.g., any minimum distance cutoff among the sampled nucleon pairs.

For each collision we draw a random impact parameter, $b$, from a distribution $dN/db \propto b$. This corresponds to the distance between the centers of the colliding ions. For each ion, we shift, then, the coordinates of the corresponding nucleons by $+b/2$ and $-b/2$, respectively. The direction of this shift, i.e., the direction of the impact parameter, defines the $x$ direction in the transverse plane. Two nucleons, belonging to different parent nuclei, interact in the transverse plane if their distance is less than
\begin{equation}
    D=\sqrt{\sigma_{\rm NN}/\pi},
\end{equation}
where $\sigma_{\rm NN}$ is the inelastic nucleon-nucleon cross section, which at top LHC energy is approximately 7 fm$^2$. A nucleon is labeled a participant if it undergoes at least one scattering with a nucleon coming from the target nucleus. The total number of participants is dubbed $N_{\rm part}$.

From this point on, we describe the collisions process and the subsequent QGP formation by means of the \trento{} model of initial conditions \cite{Moreland:2014oya}. We center on top of each participant nucleon a two-dimensional Gaussian distribution of participant matter of width $\omega=0.5$ fm. The sum of these participant-level distributions gives the so-called thickness functions of the colliding nuclei. For, say, nucleus $A$, the thickness reads:
\begin{equation}
    t_A({\bf x}) = \sum_{i=1}^{N_{\rm part,A}} \frac{\lambda_i}{2 \pi \omega^2} e^{-\frac{({\bf x-{\bf x}_i})^2}{2\omega^2}} ,
\end{equation}
where ${\bf x}_i$ is the location of the $i$th participant inside nucleus $A$, and $\lambda_i$ is a normalization drawn independently for each participant from a gamma distribution of unit mean and standard deviation equal to $1/\sqrt{2}$. This particular tuning of the parameters $\omega$ and $\lambda$ proves powerful in phenomenological applications at LHC energies \cite{Giacalone:2017dud}. For the collision of nucleus $A$ against nucleus $B$, the entropy density in the transverse plane of the QGP created in the collision process is then given by:
\begin{equation}
    s({\bf x}) \propto \sqrt{t_A({\bf x})t_B({\bf x})}.
\end{equation}
From the knowledge of the entropy density, we can obtain all the quantities necessary to draw Fig.~\ref{fig:4}. 

We first need the total entropy in each event:
\begin{equation}
    S \propto \int_{\bf x} s({\bf x}).
\end{equation}
The entropy, $S$, is used to sort events in centrality classes. As discussed below, experiments are unable to determine the impact parameter of the collisions, therefore, they rely on auxiliary variables to reconstruct it \cite{Yousefnia:2021cup}. These variables are of more or less direct variants of the number of particles ($N$ in Eq.~(\ref{eq:mpt})) produced in a given event, which is, in turn, in a nearly one-to-one correspondence with the entropy of the QGP \cite{Giacalone:2020ymy}. The distribution of $S$ provides, hence, a means to sort the simulated events into centrality classes in a way which is consistent with the methodology of the experimental collaborations. The centrality is in general well-approximated by the relation
\begin{equation}
    c = \frac{\pi b^2}{\sigma_{\rm inel}},
\end{equation}
where $b$ is the impact parameter and $\sigma_{\rm inel}$ is the inelastic nucleus-nucleus cross section, $\sigma_{\rm inel}\approx 770$ fm$^2$ in \pbpb{} collisions, and $\sigma_{\rm inel} \approx 570$ fm$^2$ in \xexe{} collisions. Expressing $c$ as a percentile fraction, one can then relate the centrality values shown in Fig.~\ref{fig:4} to the impact parameters, irrespective of the specific variable used to define the percentile.

In each centrality class, we evaluate the Pearson correlator of Eq.~(\ref{eq:5}), i.e.,
\begin{equation}
\label{eq:M5}
    \rho(v_2^2,\bra p_t \ket) = \frac{\bigl  \bra \delta v_2^2 \delta \bra p_t \ket \bigr \ket}{\sqrt{\bigl \bra \bigl (\delta v_2^2 \bigr)^2 \bigr \ket  \bigl \bra \bigl (\delta \bra p_t \ket \bigr)^2  \bigr \ket}}.
\end{equation}
To do so, we need the knowledge of $v_2$ and $\bra p_t \ket$ in each event. To exhibit results that have a statistical uncertainty comparable to that of the experimental ATLAS measurements, we follow recent theoretical developments \cite{Bozek:2020drh,Giacalone:2020dln,Schenke:2020uqq}, and evaluate the Pearson coefficient by means of accurate initial-state predictors. We consider that, in a given centrality class, the value of $v_2$ in the final state is linearly correlated with the value of the elliptic anisotropy, $\varepsilon_2$, of the initial state of the QGP, corresponding to the normalized quadrupole moment of the entropy density \cite{Teaney:2010vd},
\begin{equation}
    \varepsilon_2 = \frac{ | \int_{\bf x} |{\bf x}|^2 e^{i2\Phi}  s({\bf x})~ | }{\int_{\bf x} |{\bf x}|^2 s({\bf x})},
\end{equation}
where $\Phi=\tan^{-1} (y/x)$. Further, we consider that the value of $\bra p_t \ket$ is linearly correlated with the value of the energy, $E$, carried by the QGP \cite{Giacalone:2020dln} after its formation. The transverse energy density of the system, $e(\bf x)$, is evaluated from the conformal equation of state of quantum chromodynamics at high temperature:
\begin{equation}
    e({\bf x}) \propto s({\bf x})^{4/3},
\end{equation}
so that the total fluid energy reads:
\begin{equation}
    E \propto \int_{\bf x} e({\bf x}).
\end{equation}
Replacing, then, $v_2$ with $\varepsilon_2$ and $\bra p_t \ket$ with $E$ in Eq.~(\ref{eq:M5}) leads to a very good approximation of the results obtained at the end of full hydrodynamic simulations \cite{Giacalone:2020dln}. Our curves in Fig.~\ref{fig:4} result from $2\times10^6$ simulations of \pbpb{} collisions, and $5\times10^6$ simulations of \xexe{} collisions (for both $\gamma=0$ and $\gamma=26.93^\circ$). The statistical errors are calculated with the jackknife resampling method. The simulations are performed by means of a code written in \texttt{Python 3} which we have developed for this application.

\bigskip
In a final note concerning the experimental data points, we emphasize that, since the impact parameter of the collisions can not be determined experimentally, the ATLAS collaboration has made use of two different variables to sort their events into centrality classes \cite{ATLAS:2019peb}, and perform the measurement of $\rho(v_2^2,\bra p_t \ket)$ \cite{ATLAS:2021kty}. The first variable is the raw number of charged particles, $N_{\rm ch}$, observed in the central region of the detector, corresponding to the pseudorapidity window $|\eta|<2.5$, where the pseudorapidity is defined by $\eta=-\ln \tan (\Theta/2)$, where $\Theta$ is the polar angle in the $(y,z)$ plane of Fig.~\ref{fig:1}. The second variable is the total transverse energy, $\sum E_T$, deposited by the products of the collisions in forward calorimeters covering $3.2<|\eta|<4.9$. Additionally, the measurement has been performed for different kinematic ranges of the hadrons used to build the $\rho(v_2^2,\bra p_t \ket)$ observable. We recall that the momentum of a given hadron in the transverse plane is denoted by $p_t$. The observable has been calculated for $0.5<p_t<2$ GeV and for $0.5<p_t<5$ GeV, both considered in Fig.~\ref{fig:4}.

\end{document}